\documentclass[9pt,twocolumn,twoside]{pnas-new}

\templatetype{pnasresearcharticle} 

\makeatletter
\fancypagestyle{firststyle}{
   \fancyfoot[R]{\footerfont PNAS\hspace{7pt}|\hspace{7pt}\textbf{\today}\hspace{7pt}|\hspace{7pt}vol. XXX\hspace{7pt}|\hspace{7pt}no. XX\hspace{7pt}|\hspace{7pt}\textbf{\thepage\textendash\pageref{LastPage}}}
   \fancyfoot[L]{\footerfont\@ifundefined{@doi}{}{\@doi}}
   \fancyhead[LO,LE,RO,RE]{} 
}
\makeatother

\begin{document}
\title{Aubry-André Localization Transition for an Active Undulator}

\author[a,1]{Christopher J. Pierce}
\author[a,1]{Tianyu Wang}
\author[a]{Dmitri Kalinin}
\author[a]{Andrew Zangwill}
\author[a,2]{Daniel I. Goldman}

\affil[a]{School of Physics, Georgia Institute of Technology, Howey Physics Bldg, 837 State St NW, Atlanta, GA 30332}

\leadauthor{Pierce}


\significancestatement{Recent work has demonstrated that undulating self-propelling systems (like elongate robots and snakes) in \textit{ordered} environments mimic aspects of diffracting and scattering electrons and photons. Here we study the dynamics of an undulating robot transiting through a one-dimensional \textit{disordered} landscape. Laboratory experiments, simulations, and a generalization of resistive force theory reveal that such a system undergoes a localization transition as the aperiodicity of the landscape increases. The transition quantitatively resembles the transition in the Aubrey-André model of electron transport in one-dimensional pseudorandom potentials; this model is an alternative to the better-known case of Anderson localization in truly randomly disordered lattices. This work demonstrates the surprising utility of applying tools from condensed matter physics to understanding locomotion in disordered environments.}

\equalauthors{\textsuperscript{1} C.J.P. contributed equally to this work with T.W.}
\correspondingauthor{\textsuperscript{2}To whom correspondence should be addressed. E-mail: author.two\@email.com}


\begin{abstract}
The transport of deformable self-propelling objects like bacteria, worms, snakes, and robots through heterogeneous environments is poorly understood.  In this paper, we use experiment, simulation, and theory to study a snake-like robot as it undulates without sensory feedback through a narrow channel containing a linear array of boulder-like hemispherical obstacles. The profile of the boulder landscape approximates a one-dimensional potential introduced by Aubry and André (AA) to study wave function localization in aperiodic lattices. The AA model provides a deterministically disordered alternative to the better-known phenomenon of Anderson localization, which occurs in truly random disordered lattices. When the boulder landscape is strictly periodic, the robot can pass completely through the channel. But if the landscape is sufficiently aperiodic, the robot becomes trapped and fails to exit the channel.  The metrics we use to quantify this transition -- including exponential distributions of robot position when localized -- agree well with earlier experimental and theoretical work on a localization transition that occurs when quantum waves interact with the AA potential. A theoretical treatment of the robot's motion using resistive force theory modified to include spatially varying drag forces reproduces the behavior we observe. Further, our results indicate that the transition is generated by large fluctuations in the driving torques required for self-propulsion. These results point to a potentially fundamental connection between classical and quantum wave mechanics and the locomotion of undulators. Our study illustrates how analogies with models from condensed matter physics and wave optics can lead to the discovery of principles of self-propulsion in non-periodic landscapes. 

\end{abstract}

\dates{This manuscript was compiled on \today}
\doi{\url{www.pnas.org/cgi/doi/10.1073/pnas.XXXXXXXXXX}}

\maketitle
\thispagestyle{firststyle}
\ifthenelse{\boolean{shortarticle}}{\ifthenelse{\boolean{singlecolumn}}{\abscontentformatted}{\abscontent}}{}

\firstpage[8]{3} 

\dropcap{E}quations of motion derived for point particles subject to simple external forces make predictions that agree well with experimental observations, {\it e.g.}, electrons drifting through a current-carrying wire, biomolecules diffusing through a gel, and asteroids hurtling through space. Compared to these systems, much less attention has been paid to systems that actively self-deform and self-propel, {\it e.g.}, bacteria that swim through fluids by rotating their flagella \cite{berg1973bacteria,lauga2016bacterial}, snakes that move through sand by bending their bodies \cite{guo2008, marvi2014sidewinding,jayne1986kinematics} and robots that use legs to walk on granular media\cite{li2009sensitive}. For locomotors of this kind, forward motion results from environmental reaction forces such as friction and fluid drag. These forces typically vary over both the length of the object and in time because they arise in response to complex, internally powered and controlled body deformations. Unlike a point particle, the principles that govern the transport of a self-propelled and self-deforming object must reflect both the geometry of its self-deformations---known as gaits---and the detailed nature of the reaction forces exerted on it by its environment. 

Despite these complications, researchers from various disciplines have developed useful descriptions of the self-propulsion of objects through environments that are homogeneous and isotropic \cite{taylor1952analysis, gray1955propulsion, berg1973bacteria, shapere1989geometry, guo2008, lauga2009hydrodynamics, Zhang2014-fi, marvi2014sidewinding,lauga2016bacterial, Rieser2024}. The same cannot be said for environments that are inhomogeneous in space, time, or both. In the living world, this is the situation faced by spermatozoa swimming in complex heterogeneous fluids,  snakes slithering through a garden, and horses competing in a steeplechase. In the human-made world, familiar examples include autonomous search-and-rescue robots~\cite{murphy2017disaster}, extraterrestrial rovers~\cite{maimone2006autonomous,shrivastava2020material}, and  agricultural robots design to survey and protect crops~\cite{bechar2016agricultural,gcr}. 

Previous work revealed that snakes, nematode worms, and an elongated limbless robot successfully traverse certain periodically heterogeneous environments by exploiting a strategy they typically use to traverse homogeneous environments. That is, they impose a quasi-sinusoidal bending wave over the length of their bodies \cite{park2008enhanced,majmudar2012experiments,rieser2019dynamics,schiebel2019mechanical, Wang2023mechanical,transeth2008snake,wang2020directional}.  In one study, snakes and an undulating robot deforming in this way passed through a one-dimensional array of vertical posts in a manner reminiscent of the way light diffracts from an array of equally spaced slits cut into an opaque screen\cite{schiebel2019mechanical, rieser2019dynamics}.  In another study, nematode worms and a limbless snake-like robot passed through a two-dimensional array of vertical posts at constant speed by mimicking the Bragg condition of crystallography and matching (approximately) their undulation wavelength to the periodicity of the array\cite{Wang2023mechanical}. 

Natural terrains are rarely periodic, and tools for understanding self-propulsion in aperiodic environments are lacking. Therefore, as a step towards identifying the principles of self-propulsion in such environments, here we exploit experiments, simulations, and theoretical calculations to study the quasi-one-dimensional locomotion of a limbless and self-propelled snake-like robot through periodic and aperiodic terrains. The robot uses sinusoidal deformations of its body--without sensory feedback--to traverse a straight, narrow channel containing a linear array of hemispherical obstacles. The undulating robot passes through the channel at constant speed when the arrangement of boulders is periodic. However, when the boulder arrangement includes a sufficiently large \textit{aperiodic} component, the robot becomes trapped along the channel. We demonstrate a striking similarity between our robot's transition from a ballistic state to a localized state and the transition made by a quantum wave from a freely propagating state to a localized state when both interact with a sufficiently aperiodic environment.  Theoretical calculations using a generalized form of resistive force theory reproduce our observations, indicating how analogies with models from condensed matter physics and wave optics can lead to discoveries of principles governing self-propulsion in non-periodic landscapes. 

\section{Results and Discussion}

\subsection*{Undulating Transport in Periodic and Aperiodic Terrains: Experiment}

Our snake-like robot was constructed from a set of $N=9$ servomotors (Dynamixel XC330-M288-T) connected by hinged joints (Fig.~1A). To mimic the undulatory gaits observed for certain snakes, nematode worms, and sandfish lizards~\cite{Rieser2024}, we imposed a ``serpenoid" ~\cite{hirose1993biologically} deformation pattern on the robot's body (Fig.~1B, C) by specifying that the joint angles vary in time as (Fig~1B) 
\begin{equation}
\begin{aligned}
\alpha_m(t) &= A\sin(2\pi  m/N - \omega t), \hspace{1cm} m = 1, \dots N
\label{eq:serpenoid}
\end{aligned}
\end{equation}
This produces an approximately sinusoidal bending wave of lateral displacement that travels from the robot's head to its tail over one period.\footnote{Because the number of joints is finite, it is necessary to multiply the factor of $2\pi m/N$ in Eq.~(1) by a factor of $(N+1)/N$.} We used $A=50^\circ$ and $\omega=0.0625$ ${\rm s}^{-1}$ for all the work reported in this paper. The robot's joints are actuated by servomotors operating under position control. Each motor includes a built-in low-level PD controller, with default proportional and derivative gains set to $K_P = 8.5$ and $K_D = 31.2$, respectively. To complement our experiments,  we developed a multibody simulation,\footnote{Using CoppeliaSim~\cite{coppeliaSim}}, which replicates the mechanical properties and control strategies of the physical robot. These simulations reproduce the behavior of the real robot well and therefore allowed us to systematically explore a broader range of parameter variation on a scale impractical for physical trials alone. As in our previous experiments~\cite{Wang2023mechanical}, we operated the robot in a regime where coasting \cite{Rieser2024} is negligible. That it, self-propulsion ceases almost immediately when the deformations cease because frictional dissipation dominates inertia \cite{hu2009mechanics}. 

We limited our robot to quasi-one-dimensional motion by confining it to a straight channel formed by two parallel rigid walls. Inside the channel, we arranged obstacles in a pattern inspired by a potential introduced by Aubry and André (AA) to study the localization of electrons in one-dimensional aperiodic lattices~\cite{aubry1980analyticity}; for a useful pedagogical discussion of the AA model, see~\cite{dominguez2019aubry}. This potential can be thought of as a pseudo-randomly disordered version of the truly randomly disordered potential used by Anderson in his pioneering study of electron localization.\cite{anderson1958absence} In the AA model, electrons interact with a base potential  $\propto \sin(kx)$ and a perturbation potential $\propto \sin(\beta kx)$.  The eigensolutions associated with the AA potential change character from propagating to localized as a function of the value of the commensurability parameter $\beta$ and the relative amplitude of the two components of the potential~\cite{Modugno2009,Albert2010}. 

\begin{figure}[!ht]
\centering
\includegraphics[width=1\linewidth]{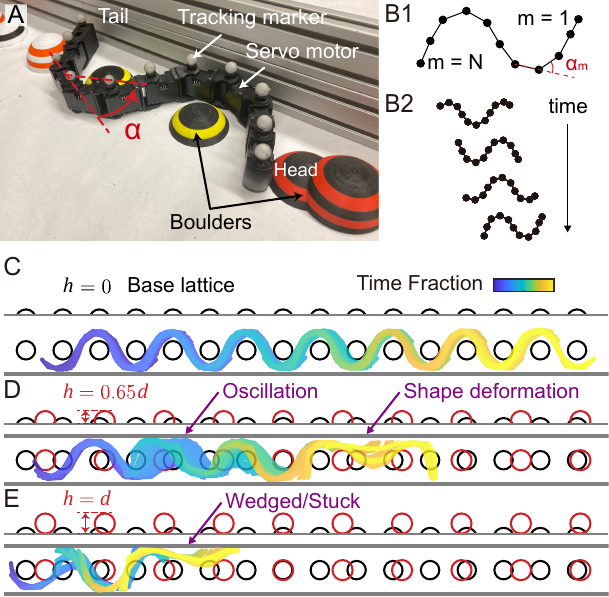}
\caption{A snake-like robot and its self-propelled motion through a narrow channel lined with rigid hemispherical boulders. (A) A robot made from $N=9$ servo-motors connected serially at hinged joints interacts with an obstacle course of 3D-printed truncated spheres (one wall of the channel has been removed for visibility); (B1) The angle $\alpha_m$ at each of $N$ hinged joints is specified as a function time to produce bending waves which move from the head to the tail; (B2) The serpenoid gait of the moving robot which results from the specification of the angles in B1; (C) $h=0$: Upper panel side view shows the periodic base lattice of boulders; lower panel top view shows ballistic motion through the channel; (D) $h=0.65d$: The robot moves forward initially, oscillates around a trapping point, escapes, and stops farther down the channel because it deforms and the motors cannot supply the torque required to return the joint angles to the values needed to maintain the serpenoid shape; (E) $h=d$: The robot which moves forward initially but almost immediately stops by wedging itself between the boulder array and the side wall.}
\end{figure}

In detail, we first placed in the channel a row of equally-spaced 3D-printed spheres (diameter $d=6.3$ cm) truncated so their tops lay a distance  $s=0.2d$ above the channel floor (top panel of Fig.~1C).\footnote{This was the smallest value of $s$ where the robot moved one body length per one cycle of the imposed deformation wave.} The separation $a$ between neighboring spheres (hereafter ``boulders") was chosen as one-half the wavelength of the deformation wave imposed on the robot. This choice facilitates the transmission of the robot through the array\cite{Wang2023mechanical}. To this periodic array of boulders,  we added a perturbation in the form of a second row of equally spaced boulders with nearest-neighbor separation $a/\beta$ and truncated so their tops lay a height $h$ above the channel floor. For convenience, we will use the term ``AA terrain" when referring to this obstacle course inside the channel.  Our experiments and simulations consisted of placing the robot at one end of the channel, initiating the gait sequence, and measuring its body shape and center-of-mass position as a function of time for different values of $\beta$ and different heights of the perturbing boulder array between $h=0$ (fully buried) to $h=d$ (fully exposed). 

We begin with $\beta=(\sqrt{5}-1)/2$. This choice of an irrational number implies that the two boulder lattices are incommensurate. Our principal observation is that the robot's dynamics change dramatically as a function of the height of the perturbing boulders. When only the base periodic array of boulders is present ($h=0$), the robot passes through the channel at constant speed, gaining thrust from reaction forces at 3 points of contact with the boulder array (lower panel of Fig.~1C). Any value $h>0$ presents the robot with an {\it aperiodic} terrain, and we find that there are intermittent periods when it moves {\it backward} in the channel. These periods of backward motion become more frequent as $h$ increases until, finally, the robot fails to exit the channel altogether. This happens in one of two ways: (i) the robot becomes dynamically {\it trapped} and oscillates around some point in the channel (Fig.~1D); or, much more frequently, (ii) we {\it terminate} the trial and localize the robot at that point because it flips out of the lattice, a servomotor overloads, it wedges itself against the channel wall, or the servomotors are otherwise unable to provide the bending torques needed to maintain the prescribed serpenoid shape (Fig.~1E), and thus fails to provide to produce forward thrust.

Fig.~2A illustrates the distance $X(t)$ the robot travels down the channel as a function of time for a typical trial for several values of $h$. The straight line for $h=0$ (black) reflects ballistic transport through the periodic base lattice. The $h=0.5d$ trial (green) begins with the robot moving forward at not quite constant speed due to the presence of short backward motion events. It then becomes trapped in an oscillatory state for some time, escapes, and then gets trapped in a second oscillatory state farther down the channel until the trial ends. The $h=0.65d$ trial (blue) begins similarly but ends abruptly by termination (horizontal dashed line). The $h=d$ trial (red) features a quick termination very near the channel entrance.

\begin{figure}[t!]
\centering
\includegraphics[width=.8\linewidth]{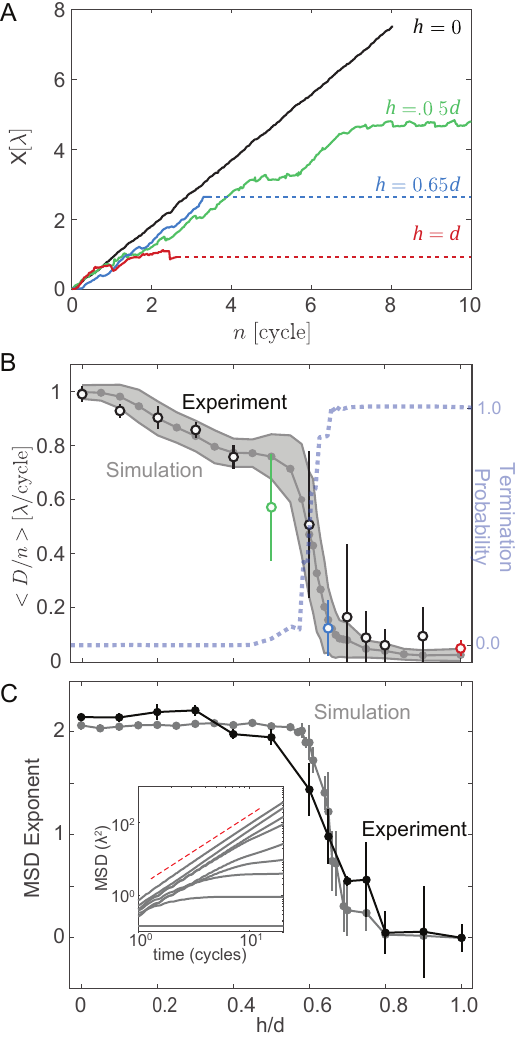}
\caption{Robot Trajectory data for $\beta=(\sqrt{5}-1)/2$. (A) Distance traveled down the channel (in units of the wavelength of the bending wave) as a function of time (in units of the period of the deformation wave) for a typical trial for four values of the perturbing lattice height $h$. (B) Left scale and data points: the average distance traveled before localization or exit from channel, divided by the number of completed deformation wave periods (in units of the undulatory wavelength). The small open circles are experimental points; the three colored circles correspond to the three colored curves in Fig.~2A. The small filled circles are from the simulation. The standard deviations are indicated by vertical lines for the experiment points and the width of the gray shaded region for the simulation. Right scale and blue dotted curve: the probability that a trial ends by termination. (C) The average exponent as determined by a power law fit to the mean squared displacement for different values of $h/d$ as a function $h/d$ for the experiment (black) and simulation (gray). Inset shows MSD curves from the simulation for several values of $h/d$; the red dashed line indicates quadratic (ballistic) scaling.}
\end{figure}

Fig.~2B summarizes the results of many experimental and simulation trials as a function of $h/d$. The left scale and the data points report the average distance $\left\langle D \right\rangle$ traveled by the robot before it exits the channel or localizes. The right scale and the dotted blue curve indicate the probability that a trial ends by termination. Beginning with the $h/d=0$ case of ballistic motion through the base lattice, the robot's mean travel distance decreases as $h/d$ increases because the frequencies of both backward motion events and transient oscillatory trapping events increase. The precipitous drop in $\left\langle D \right\rangle$ around $h/d\approx0.6$ signals that a dramatic change in the dynamics has occurred. After this transition, the robot almost always localizes by termination soon after it enters the channel. 

For many particle transport problems, it is possible to identify dynamical transitions from the fact that the mean square displacement (MSD) varies at long times as $\langle X^2(t)\rangle \sim at^b$. For example, Mokhtari and Zippelius\cite{Zippelius2019} simulated the motion of active filaments through a porous medium and discovered a transition from ballistic ($b=2$) to diffusive ($b=1$) to localized ($b=0$) behavior as a function of the stiffness of the filament. In a similar way, the MSD exponent we calculate from our robot data (experiment and simulation) demonstrates that localization replaces ballistic motion when $h/d$ exceed a critical value (Fig.~2C).

Statistical distributions of the total travel distance $D$ provide another quantitative measure of transport change. We plot these distributions as solid curves in Fig.~3A for the same values of $h/d$ used in Fig.~2A. This illustrates the transport behavior before (green), during (blue), and after (red) the transition. Fitting these distributions to the function $Ae^{-(D-D_0)^{\alpha}}$ (dashed curves in Fig.~3A) reveals a transition in the distribution shape from Gaussian ($\alpha = 2$) to exponential ($\alpha = 1$) (Fig.~3B). Remarkably, curves very similar to Fig.~2B and 3B appear in the work of Roati and
co-workers\cite{Roati2008anderson,AspectInguscio2009} who followed the motion of a Bose-Einstein condensate of ${\rm ^{39}K}$ atoms through a one-dimensional optical potential of the AA type and observed the predicted transition from a de-localized quantum state to a localized quantum state.
 
To provide further evidence that a robot undulating through an AA terrain behaves very much like a wave propagating through an AA potential, we next develop a theoretical model that accounts well for the results of our experiments and simulations.

\begin{figure}[t!]
\centering
\includegraphics[width=.8\linewidth]
{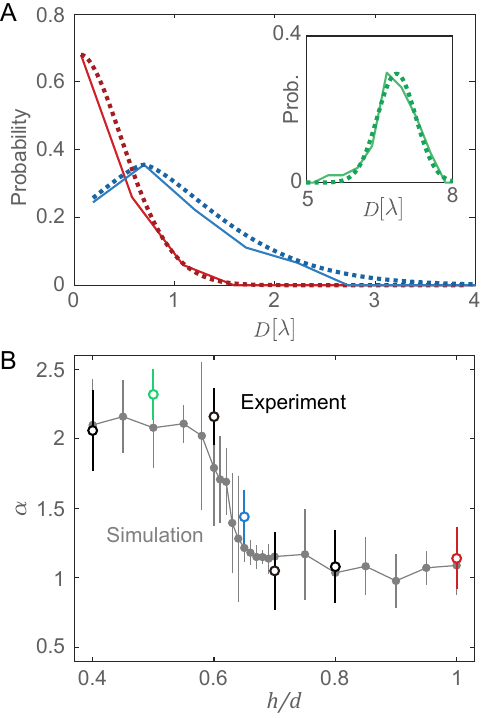}
\caption{Distributions of robot travel distance. (A) Solid curves are the experimental data for $h=0.5d$ (green), $h=0.65d$ (blue), and $h=d$ (red); dashed curves are fits to the stretched exponential function $A\exp{[-(D-D_0)^{\alpha}]}$. (B) The variation of the fitting parameter $\alpha$ with $h/d$. Open circles are experimental results; closed circles are simulation results. Error bars represent 60\% confidence intervals.}
\end{figure}

\subsection*{Undulating Transport in Periodic and Aperiodic Terrains: Theory}

No single theory models the dynamics of an extended self-deforming object in all circumstances. However, when such an object moves through a homogeneous overdamped environment, so-called resistive force theory (RFT) \cite{gray1955propulsion,Zhang2014-fi} provides a good description. In this section, we describe a generalization of this theory capable of describing the undulatory transport of a snake-like robot through an inhomogeneous environment.   

Following Gray and Hancock~\cite{gray1955propulsion}, we replace the robot by a flexible one-dimensional filament of length $L$ on the $x$-axis. We model the effect of the robot's internally generated bending wave by specifying the transverse displacements of the filament: 
 \begin{equation}
    y(x,t)=b\sin\left(2\pi[x/\lambda-t/T]\right).
 \end{equation}

\noindent 
The center-of-mass of the filament (hereafter ``undulator") moves along the $x$ axis because the bending described by Eq.~(2) induces time-varying contact and resistive forces. The theory is designed for overdamped, low-coasting situations where inertia can be neglected. Accordingly, Gray and Hancock derived an expression for the undulator's center-of-mass velocity, $V_{\rm cm}$, by setting to zero the sum of all the forces that act upon it. Their expression has been used with success to describe the transport of real undulators over flat frictional surfaces\cite{rieser2019dynamics} and through viscous fluids\cite{gray1955propulsion} and frictional fluids like sand\cite{Zhang2014-fi}. 
 
The key parameter in RFT is a single number called the drag anisotropy. This quantity, $\xi=f_\perp/f_\parallel$, is the ratio of the components of the total force that act perpendicular and parallel to the surface of the undulator. When $\xi>1$, the center-of-mass moves in the direction opposite to the direction of the propagating bending wave. This is the case for snakes on sand\cite{schiebel2020mitigating,rieser2021functional}, limbless robots with passive wheels on hard ground\cite{rieser2019dynamics}, and nematode worms\cite{rabets2014direct} or sperm cells in low Reynolds number fluids\cite{gray1955propulsion}, By contrast, the center-of-mass moves in the same direction as the deformation wave for systems with $\xi<1$ like centipedes on water surfaces\cite{diaz2022water} and some marine worms in bulk fluids \cite{taylor1952analysis}. 

A simple way to generalize RFT for use with inhomogeneous terrains is to permit the drag anisotropy $\xi$ to vary as a function of position.\footnote{The numerical studies of limbless locomotion by Zhang and co-workers~\cite{zhang2021friction} introduce a spatially-varying planar friction coefficient to study the effects of terrain inhomogeneity.}
An important feature of this approach is that the same body deformation wave can induce either forward or backward motion depending on the geometry of the contact between the undulator and the terrain. To illustrate how this can happen, the side and top views in Fig.~4A show the normal ($N$), frictional ($F$), and gravitational ($mg$) forces present when a thin rod strikes a hemispherical boulder tangentially. The magnitude and direction of the net force $\vec{f}$ on the rod depend sensitively on the exact point of contact with the boulder. For example, when the rod is moving to the right, it is straightforward to confirm that $\xi>1$ on the left side of the boulder (moving ``uphill") while $\xi <1$ on the right side of the boulder (moving ``downhill"). 

We have not attempted to calculate the spatial variation of the drag anisotropy for any actual terrain. Instead, we construct a minimal model that uses simple analytic anisotropy functions $\xi(x)$ that possess the same spatial periodicities as the AA terrain. With $\xi(x)$ in hand, the generalization of the expression for $V_{\rm cm}$ derived in Ref.~\cite{gray1955propulsion} for small amplitude undulations can be shown to be (see the Supplemental Material)

\begin{equation}
    V_{\rm cm}=\frac{4\pi^2 b^2}{L\lambda T}\int_{-L/2}^{L/2} dx\left[\xi(X_{\rm cm}+x)-1\right]\cos^2{[2\pi(x/\lambda-t/T)]},
        \end{equation}
        
\bigskip
\noindent
where $X_{\rm cm }(t)$ is the position of the center-of-mass of the undulator in the frame of laboratory.  

The top panel of Fig.~4B shows the spatial variation of the total force (arrows) exerted on the undulator (solid curve) when $\xi(x) = \Delta\sin(kx) + 1$ (dashed curve). These forces sum to zero as required. However, when $\Delta$ is large enough, they provide sufficient forward thrust locally to propel the robot in that direction at speed $V_{\rm cm}=\lambda/T$ (see the Supplemental Material). This is consistent with the ballistic transport seen in Fig.~2A for the $h=0$ simple periodic boulder field.

\begin{figure}[t!]
\centering
\includegraphics[width=.8\linewidth]{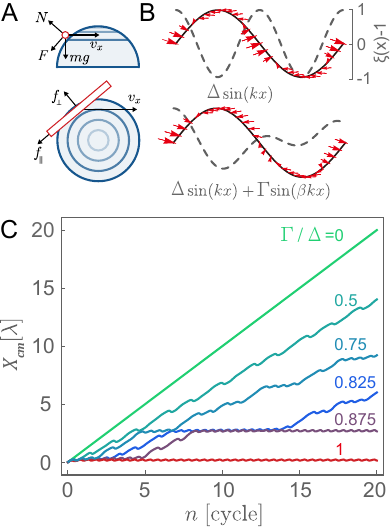}
\caption{Resistive Force Theory modified to include a spatially varying drag anisotropy. (A) Side (upper panel) and top (lower panel) views of a thin rod making contact with a curved frictional surface. See text for discussion of the forces shown. (B) The net force pattern (red) associated with a periodic drag landscape (upper panel) always produces net thrust to the right, while the force pattern associated with an aperiodic drag landscape (lower panel) can yield zero net motion or alternating leftward and rightward motion. (C) RFT model results for the undulator's center-of-mass position. The motion is ballistic when $\Gamma/\Delta=0$ (green curve). But as $\Gamma/\Delta$ increases, the undulator exhibits increasingly frequent bouts of backward motion and transient trapping until, above a critical value, it becomes permanently trapped and oscillates indefinitely around a single point (red curve).}
\end{figure}

A drag anisotropy function which mimics the AA terrain is  
\begin{equation}
    \xi(x) = \Delta\sin(\pi x/a) + \Gamma \sin(\beta \pi x/a)+1. 
\end{equation}
This function is aperiodic when $\beta=(\sqrt{5}-1)/2$ and the distribution of forces it produces (lower panel of Fig.~4B) leads to time-varying values of $V_{\rm cm}$ that can be positive or negative. Fig.~4C shows results using Eq.~(4) for the distance traveled by the undulator as a function of time. As $\Gamma$ increases, the transport evolves from ballistic, to near-ballistic (due to the presence of backward motion events), to near-ballistic punctuated by episodes of transient oscillatory trapping, to a final oscillatory trapped state. The similarity of these theoretical results to the experimental/simulation results of Fig.~2A is striking. By construction, the undulator in our one-dimensional RFT calculations cannot assume any of the configurations that led to localization by termination for our robot. However, we demonstrate below that the force fields that produce oscillatory trapping in the RFT model are similar to the force fields that trigger termination in our experiments and simulations.  

Fig.~5A plots our RFT results for the average distance traveled by the undulator as a function of $\Gamma/\Delta$ for three values of $\beta$. The good qualitative agreement between the $\beta=(\sqrt{5}-1)/2$ curve (red) and the corresponding experimental data plotted in Fig.~5B (repeated from Fig.~2B) gives us further confidence that RFT captures the essential dynamics of the real robot moving through an AA terrain. The RFT results shown in Fig.~5A for rational values of $\beta$ are surprising (initially) because one reflects ballistic motion ($\beta=1/2$ in green) and the other exhibits localization ($\beta =2/3$ in blue). However, comparison with Fig~5B shows that the robot behaves in exactly the same way. Moreover, the qualitative difference in transport behavior associated with these two rational values of $\beta$ is precisely what is seen in detailed numerical studies of the AA model where the eigenstates of the Schr\"odinger equation change from delocalized to localized even when $\beta=p/q$ as long as $p\neq 1$ and $q$ are integers\cite{Modugno2009}.  

\begin{figure}[t!]
\centering
\includegraphics[width=.8\linewidth]{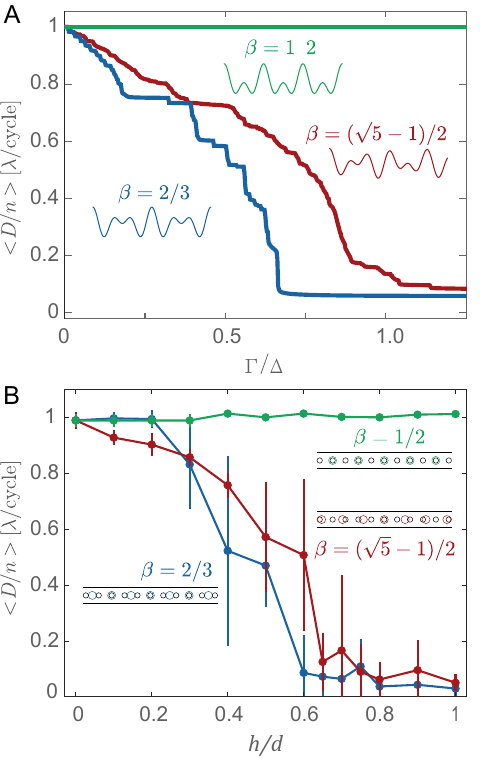}
\caption{Average distance traveled for different values of $\beta$. (A) RFT values of $\left\langle D \right\rangle$ divided by the number of completed deformation wave periods (in units of the undulatory wavelength) as a function of $\Gamma/\Delta$. (B) Experimental values of $\left\langle D\right\rangle$ as above as a function of $h/d$. Error bars indicate the standard deviations.}
\end{figure}

\section*{Trapping versus Termination}

Our robot experiments, our multi-body simulations, and our generalized RFT calculations agree with respect to the commensurability and aperiodicity conditions that determine whether an elongated undulator traverses a quasi-one-dimensional AA terrain or stops and localizes at a point within it. More precisely, they agree if ``localization" is understood to mean both oscillatory trapping (for our generalized resistive force theory) and trial termination (for our experiments and simulations). The task of this section is to show that these outcomes are indeed essentially equivalent for our purposes by describing the underlying torque patterns that lead to either case.  

A key point is that locomotion in the non-inertial, low-coasting situations studied here occurs only because the locomotor repeatedly generates and propagates a bending wave of distortion down its body. The robot does this by using its servomotors to apply the torques required to impose the joint angles specified by Eq.~(1). RFT does this by imposing a gait on the undulator defined by Eq.~(2), and the theory can be used to calculate the torques that would be required to maintain this gait. It is instructive to compare these two because it is torque-induced stress that leads the undulator and the robot to reverse, and eventually localize by trapping and termination, respectively.

We focus on $\beta=(\sqrt{5}-1)/2$ and consider first the RFT torques. Fig.~6A shows the time variation of the torque near the head of the undulator for three values of $\Gamma/\Delta$. When $\Gamma/\Delta=0$ (top panel), the terrain is periodic, the undulator motion is ballistic, and the torque oscillates with the same period as gait of the undulator. This occurs because of the matched spatial periodicity of the robot's gait and the terrain. Over the duration of a cycle, the robot returns to the same relative position in the lattice; hence the temporal periods of the torque wave and the gait itself are the same. 
\begin{SCfigure*}[\sidecaptionrelwidth][t!]
\centering
\includegraphics[width=0.8\textwidth]{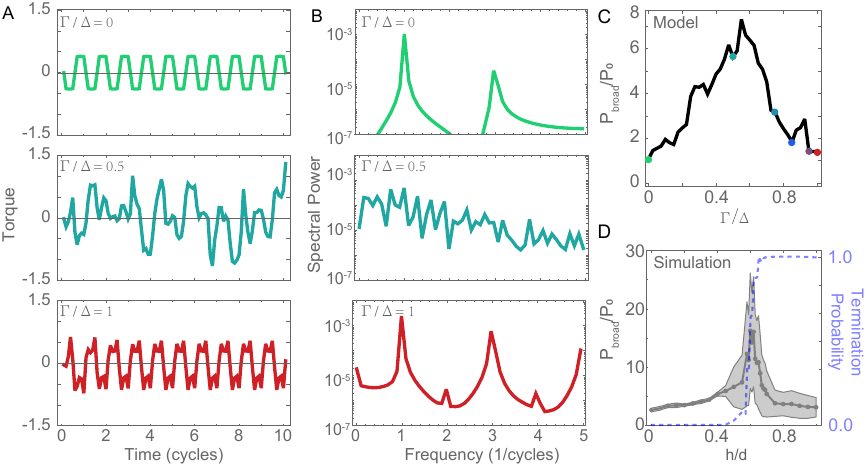}
\caption{ Torque exerted by an undulator in $\beta=(\sqrt{5}-1)/2$ AA terrain at its head. (A) RFT results for the torque as a function of time for three values of $\Gamma/\Delta$. (B) Power spectrum of the torque series in (A). (C) Relative spectral power $P_{\rm broad}/P_{\rm 0}$ calculated from (B) as a function of $\Gamma/\Delta$. Filled circles correspond with the curves shown in Fig. 4C. See text for discussion. (D) Simulation results (filled circles) for $P_{\rm broad}/P_{\rm 0}$. Shaded region indicates the standard deviation. Blue dashed curve is the probability that a simulation trial ends by termination (reproduced from Fig 2B).}
\end{SCfigure*}

When $\Gamma/\Delta = 0.5$ (Fig.~6A, middle panel), the incommensurate perturbing component of the AA terrain induces transient episodes of both backward motion and oscillatory trapping. In breaking the spatial periodicity match between the gait and the terrain, this in turn produces a torque that varies aperiodically in time. Finally, after the magnitude of the perturbation has increased to the point that $\Gamma/\Delta =1$ (Fig.~6A, bottom panel), the undulator localizes and the torque again varies periodically in time at the oscillatory trapping frequency, which is matched to the gait frequency. As in the case of ballistic motion at $\Gamma/\Delta =1$, the trapped state produces motion where the robot returns to the same relative position after a cycle, but instead of advancing by one boulder spacing, the undulator returns back to the same point where it began within the trap.

Our aim is to show that, in terms of the torque patterns, the RFT undulator and the robot transition similarly to a localized state, despite differences between the behaviors of oscillatory trapped states and localization by termination. To that end, we next Fourier transform the time series in Fig.~6A to produce the power spectra in Fig.~6B. The peaks in the spectral curves in the top and bottom panels of Fig.~6B identify the robot's gait frequency and the trapping frequency. The broad-band noise in the middle panel of Fig.~6B is present because, when $\Gamma/\Delta = 0.5$, the AA terrain is sufficiently aperiodic that the undulator encounters many different types of local roughness, each inducing a short bout of reversed motion, and each requiring a different torque pattern to maintain the prescribed gait. As $\Gamma/\Delta$ increases, these short bouts of reversal occur more and more frequently, eventually stringing together in series to produce longer and longer periods of oscillatory trapping at fewer and fewer distinct locations (See Fig. 4C). Because the aperiodic torque components are associated with transitions between reversals/traps and forward motion, when the undulator becomes permanently trapped, the power spectrum again becomes dominated by peaks at the undulation frequency (Fig. 6B, bottom panel). 

Fig.~6C further illustrates this scenario by plotting the ratio $P_{\rm broad}/P_{\rm 0}$ for the undulator as a function of $\Gamma/\Delta$. The numerator is the integrated spectral power of the broad-band torque noise. The denominator is the noise power in a narrow band centered at the gait frequency.\footnote{We calculated $P_{\rm broad}$ by integrating the power over a band of from 0-5 in units of the undulation frequency. We calculated $P_{\rm 0}$ by integrating the power near the undulation frequency (over a band from$\approx 0.9-1.1 $).} This normalized power increases at first because the number of isolated, torque-driven reversals increases as the landscape aperiodicity increases. The number of isolated reversals eventually reaches a maximum and begins decreasing when the reversals begin to string together, forming oscillatory traps, as indicated above. The noise then decreases rapidly as $\Gamma/\Delta$ increases because of the increasing coalescence of reversal events. Finally, the torque noise returns to its initial value when the undulator achieves its final state of oscillatory localization.

 Fig.~6D plots $P_{\rm broad}/P_{\rm 0}$ for the robot calculated similarly using simulation data along with the probability that a trial ends in termination (redrawn from Fig.~2B). Once again, the torque noise rises and falls as a function of the terrain aperiodicity, albeit more steeply in Fig.~6D than in Fig.~6C. The reason for this, we posit, is that the three-dimensional robot can respond to aperiodicity in many ways while the RFT undulator, which is constrained by construction to one-dimension, can do so only by reversals and oscillatory trapping. For example, when $h/d$ is small (below the peak in Fig.~6D), the robot can negotiate the terrain using small out-of-plane motions which produce less broad-band torque noise than a reversal event. As the aperiodicity becomes higher, eventually the robot fails to respond with small out-of-plane motion, and begins to pick up noise in the torque signal.
 When $h/d$ is large (above the peak), the robot can flip over, jam against a sidewall, or distort into a shape very far from serpenoid. All of the latter events trigger termination, which leads us to conclude that termination simply preempts localization by trapping and likely occurs in response to the same torque patterns. That is why, despite the somewhat \textit{\it ad hoc} rules we developed to initiate termination, the systematics of the transition to localization exhibited by our experiments and simulations qualitatively agree with both our RFT calculations and with the behavior expected from the quantum AA model. 
 
 In summary, localization via oscillatory trapping (in RFT) or by trial-terminating stopping events (in the simulation and experiments), both lead to the same basic outcome -- forward motion stops-- despite appearing qualitatively different. However, these seemingly different mechanisms of halting forward progress arise from similar underlying torque patterns, which reflect the geometry of both the undulatory gait and the landscape. 

\section*{Summary and Conclusions}
 This paper describes experiments and multi-body simulations designed to study the transport of an undulating snake-like robot through a straight and narrow channel lined with a linear array of 3D-printed boulder-type obstacles. The boulder arrangement was chosen so the terrain landscape resembled a potential used by Aubry and Andr\'e to study the localization of electrons in a one-dimensional crystal. The AA potential superposes two periodic potentials, so we studied the transport as a function of the landscape's commensurability (by tuning the ratio of the periods) and its aperiodicity (by tuning the ratio of the amplitudes)

 The robot traversed the channel at constant speed when the landscape was periodic and its periodicity matched the wavelength of the robot's bending wave. Depending on the commensurability of the landscape, increasing the aperiodicity of the landscape led to increasingly labored transport due to the increasing appearance of episodes where the robot moved backward or transiently oscillated back and forth around a fixed point. Eventually, the robot failed to exit the channel and became localized at a fixed point inside. Most often, this happened because it responded to the landscape roughness by flipping over, jamming against a channel sidewall, or greatly distorting its body shape. In all these cases, the net outcome is a lack of continued forward movement through the channel, leading to trial termination.
 
 We quantified the transition from ballistic motion to localization using the mean square displacement of the robot down the channel and the distribution of its total travel distance. The shape of the latter changed smoothly from Gaussian to exponential as we increased the aperiodicity of the terrain, very much like what was reported in an experimental study of the transport of a Bose-Einstein condensate through an optical lattice of the AA variety. Changing the commensurability of the terrain landscape produced changes in the robot's behavior in good qualitative accord with expectations based on numerical studies of the localization (or not) of quantum waves interacting with an AA potential. 
  
We reproduced most of our experimental and simulation results using a generalization of resistive force theory. Conventional RFT applies to a filamentary undulator moving in a straight line through a homogeneous environment where drag dominates inertia and a single parameter characterizes the drag. To model the heterogeneity of our terrain, we replaced the drag parameter with a drag function equal to the AA potential function. Localization in this strictly one-dimensional theory corresponds to a state where the undulator oscillates back and forth in the immediate vicinity of a single point. 

Despite its simplifying assumptions, our version of RFT produced qualitatively similar trajectories and mean travel distances to those found in our experiments and simulations as a function of the aperiodicity and commensurability of the landscape (see also supplemental videos 1-3). This good correspondence suggested that localization by oscillatory trapping and localization by termination were closely related. We investigated this by comparing the torque exerted on the head of the moving robot to the torque on the head of the RFT undulator that would be required to maintain its prescribed gait. In both cases, broad-band noise associated with transient episodes of reversal and trapping increased, peaked, and then decreased rapidly as the landscape aperiodicity increased, reaching its smallest value when the robot/undulator finally localized. From this, we concluded that the robot and the undulator behaved essentially identically except that the three-dimensionality of the robot allowed it to respond to increasing terrain aperiodicity (roughness) with termination-triggering events not available to the one-dimensional RFT undulator. In other words, localization by termination simply preempted localization by trapping. 

The similarity of our results to the cold atom experiments deserves further comment. We are aware that a microscopic quantum particle like an atom and a macroscopic self-deforming object like our robot have nothing in common except that quantum mechanics requires the atom to possess a de Broglie wave, and the act of moving requires our robot to possess a deformation wave. That being said, it is interesting to note \cite{Albert2010} that a transition from a delocalized state to a localized state is possible for both classical and quantum particles when they interact in one dimension with the AA potential. Moreover, not unlike our RFT undulator, localization of these particles takes the form of trapping by the wells of the AA potential. The same is not true for the one-dimensional Anderson model~\cite{anderson1958absence}, where the corresponding potential is random and delocalization can occur for a classical particle but never for a quantum particle. 

In closing, we suggest it is worth considering our present results in light of earlier experiments where we observed diffraction-like behavior when snakes and limbless robots passed through a one-dimensional array of posts. We also note work by Zhang and co-workers\cite{zhang2021friction} who found lensing, refraction, and wave-guiding behavior in numerical studies of a filamentary undulator sliding along a heterogeneous surface tailored to mimic optical elements. Taken together, it is tempting to speculate that there is some previously unsuspected connection between the waves associated with self-deforming mechanical systems and the more familiar waves studied in classical and quantum mechanics.

\matmethods{

\subsection*{Design and Construction of 3D-Printed Boulder Arrays for Robot Experiments}

The boulders were made from polylactic acid (PLA) and fabricated on Bambu Labs X1-E and X1-Carbon 3D printers. Each boulder was initially designed as a sphere with a diameter of 63.5 mm ($d$) and then truncated at various heights ($h$), with values ranging from 6.35 mm (0.1 d) to 63.5 mm (1.0 d) to generate different obstacle profiles. To create clear visual distinctions of contouring from a top-down perspective, the PLA filament color was changed in increments of 3.175 mm along the height of each boulder (see Fig. 1).

Each boulder included a rectangular cutout within its base (25.4 mm $\times$ 25.4 mm $\times$ 9.525 mm with a 1.59 mm corner radius) housing two rectangular magnets. These magnets allowed the boulders to be rigidly affixed to the underlying magnetic surface. For boulders where $\Delta > 50.8$ mm $(0.8D)$, an additional cylindrical extrusion was incorporated at the base, ensuring sufficient space for embedding the magnets. However, given the significant truncation height, the robot was not expected to interact with this additional extrusion, effectively making these obstacles functionally equivalent to those simulated in virtual models. To secure the magnets, four small flanges were integrated into the boulder design, allowing for an acrylic laser-cut cover. However, for boulders with high $h$ values, the covers were omitted, as the forces exerted by the robot when manipulating these obstacles were large enough to potentially dislodge the covers.

The channel used to contain and guide the boulders was constructed using 80-20 aluminum extrusions, specifically eight pieces of 25.4 mm $\times$ 50.8 mm $\times$ 1.22 m sections. Each side of the channel measured 2.44 m in length and 101.6 mm in height. The width of the channel was set at 127 mm, precisely twice the diameter of the boulders, to allow sufficient movement and interaction space. Structural integrity was maintained by connecting the four aluminum sections on each side using a centrally placed square bracket and two additional brackets at each end. The entire channel was firmly secured to the Optical Table using two 25.4 mm L-brackets positioned on opposite sides.

\subsection*{Simulations}
\subsubsection*{Simulation Framework}
To complement our experimental investigation of undulatory transport in structured environments, we developed a multibody dynamics simulation using the CoppeliaSim platform~\cite{coppeliaSim}. The simulation replicates the kinematics of the undulating robot by prescribing a traveling wave of joint angles consistent with a serpenoid gait. While the number of body segments in the simulation was chosen to support computational efficiency and broader parametric exploration, the wavelength of body undulation was kept consistent with that used in physical experiments to preserve comparable locomotor-environment interaction dynamics.

CoppeliaSim was used in conjunction with its Bullet physics~\cite{pybullet} backend. Contact interactions were resolved using Bullet’s default sequential impulse constraint solver, an iterative method that applies impulses to enforce non-penetration and friction constraints. This constraint-based formulation enables stable simulation of rigid body interactions in friction-dominated regimes, even in highly cluttered environments.

Each body segment was modeled as a rigid link with finite mass and moment of inertia, connected to adjacent segments by revolute joints permitting planar undulation. The simulated robot was driven in open loop by sinusoidal joint actuation, and ground contact was modeled using discrete point contacts with isotropic Coulomb friction. Environmental obstacles (boulders) and channel walls were modeled as rigid, immobile objects.

The simulation enabled repeatable and systematic exploration of robot-environment interactions across a wide range of disorder conditions, including multiple realizations of aperiodic landscapes and variation in boulder geometries that would be time-prohibitive to perform physically.

\subsubsection*{Robot and Environment Modeling}

The robot in simulation was constructed as a planar multibody system using the exact geometry of the physical robot, imported directly from computer-aided design models built in SolidWorks. This ensured that segment dimensions, mass distribution, and contact surface geometry were faithfully preserved, enabling direct comparison between simulated and experimental results. Adjacent segments were connected via revolute joints, allowing relative yaw motion. Joint actuation followed a serpenoid waveform of the form:
\begin{equation}
\begin{aligned}
\alpha_m(t) &= A\sin(2\pi  m/N - \omega t), \hspace{1cm} m = 1, \dots N
\label{eq:serpenoid}
\end{aligned}
\end{equation}
where $\alpha_m$ is the angle of joint $m$, $A$ is the angular amplitude, and $\omega$ is the undulation frequency. This waveform was applied directly to the joint motors under position control.

The robot interacted with a rigid planar ground surface and static environmental obstacles, including channel walls and boulders. Friction was modeled using Bullet’s default isotropic Coulomb model, which constrains the tangential contact force according to:
\begin{equation}
\|\mathbf{F}_t\| \leq \mu \mathbf{F}_n,
\end{equation}
where $\mu$ is the coefficient of friction, and $\mathbf{F}_n$ is the normal force at the contact point. Friction coefficients were tuned to approximate the high-dissipation, low-inertia regime observed in experiments, ensuring that robot motion arose primarily from body-environment interactions and ceased promptly when actuation stopped. No restitution or velocity-dependent friction was used.

Boulder layouts in simulation matched the bichromatic spatial patterns used in physical trials, including both periodic and aperiodic configurations. Because both robot and terrain geometry were maintained identically to those in experiment, contact interactions preserved the same spatial tolerances and surface features, supporting direct comparison across varying levels of environmental disorder.

\subsubsection*{Simulation Parameters and Protocol}

Simulations were conducted with a fixed time step of $\Delta t = 10$\,ms using Bullet's default constraint-based solver within the CoppeliaSim environment. Joint motors operated under position control using a proportional controller with gain $K_p = 1$. A maximum torque limit of 20\,Nm was imposed at each joint, beyond which actuation would saturate. No additional joint velocity or acceleration constraints were enforced.

The prescribed joint waveform had an undulation frequency of 0.5\,Hz, corresponding to a 2-second cycle period. Each simulation was run for a maximum duration of 10 gait cycles. The robot's initial configuration was randomized by assigning a uniformly sampled initial phase offset to the serpenoid waveform and randomizing its initial body position within the channel while remaining fully in-plane and in contact with the ground.

Simulations were terminated early if the robot flipped out of the plane, defined as any point on the robot body exceeding a fixed vertical threshold (15 cm, $~\sim0.9$ width of the body wave) in the $z$-direction. No restitution or damping was applied, allowing motion to cease naturally under frictional contact.

For each simulation, data were recorded at 500\,Hz. The following quantities were saved at each timestep: 3D position of every body point, joint angles, and joint torques. These outputs enabled computation of quantities such as displacement, velocity, and torque patterns across varying terrain configurations. To ensure statistical robustness, 1000 independent simulations were conducted for each terrain configuration.

A summary of key simulation parameters is provided in Table~\ref{tab:sim_params}.

\begin{table}[h!]
\centering
\caption{\textbf{Summary of simulation parameters.}}
\label{tab:sim_params}
\resizebox{\linewidth}{!}{
\begin{tabular}{lll}
\textbf{Parameter} & \textbf{Value} & \textbf{Notes} \\
\hline
Simulation time step ($\Delta t$) & 10\,ms & Fixed integration step \\
Motor control type & PD & Matched to physical robot \\
Joint torque limit & 20\,Nm & Saturation enforced \\
Undulation frequency & 0.5\,Hz & 2 seconds per cycle \\
Initial waveform phase & Random & Uniform over $[0, 2\pi]$ \\
Initial position & Random & Within channel, in-plane \\
Data recording rate & 500\,Hz & Consistent across all runs \\
Recorded outputs & Position, angle, torque & For all body links and joints \\
\hline
\end{tabular}}
\end{table}

\subsection*{Model Derivation}

Following Gray and Hancock, the viscous resistive force density ($\frac{dF}{dx}$) of a slender undulator of length $L$ with lateral displacement $y(x,t)$ is given by

\begin{equation}\label{forceDensity}
    \frac{dF}{dx}=(C_N-C_L)\frac{dy}{dx}\frac{dy}{dt}-C_L \frac{dX}{dt}.
\end{equation}

\noindent The variable $x\in[0,L]$ describes the position along the body in a coordinate system that moves with a point at the tail, while the variable $X(x,t)$ describes the position of a point $x$ along the body in a fixed reference frame as it moves. The variables $C_N$ and $C_L$ refer to the normal and longitudinal resistance coefficients (in this case, viscous drag coefficients), respectively, which were assumed by Gray and Hancock to be constants intrinsic to the material constituting the body and the environment. In this manuscript, we will eventually allow these constants to vary in space (along $x'$), to describe locomotion in heterogeneous environments. We note that this relation is valid in the limit of small displacements $y$. 

For sinusoidal undulations moving to the left along the x-axis,

\begin{equation}\label{BodyWaveEqn}
    y(x,t) = y_0 \sin{2\pi(x/\lambda + \omega t)},
\end{equation}

\noindent where $y_0$ is the amplitude of the body displacement, $\lambda$ is the wavelength and $\omega$ is the frequency of undulation. For simplicity, we proceed by setting $\lambda=1$ and $\omega = 1$

Substituting (\ref{BodyWaveEqn}) into (\ref{forceDensity}) yields

\begin{equation}\label{eq3}
    \frac{dF}{dx}=4\pi^2y_0^2(C_N-C_L)\cos^2{2\pi(x+t)}  -C_L \frac{dX}{dt}
\end{equation}

In the overdamped limit, the net force $\int dF$ along the body must always equal zero. Integrating both sides of \ref{eq3} with respect to x yields the following relation (for a body of unit length)

\begin{equation}\label{odeInt}
    \int_0^1 \frac{dX}{dt}dx=4\pi^2 y_0^2\int_0^1\frac{C_N-C_L}{C_L}~\cos^2{2\pi(x+t)}~dx
\end{equation}

We assume that each segment of the body moves along the X-axis at the same speed (the body does not deform along the swimming direction, but only lateral to the swimming direction), such that the integrand on the lhs of (\ref{odeInt}) is a constant. We now introduce the spatial dependence of the resistance coefficients by writing them as follows

\begin{equation}\label{dragFunc}
    \frac{C_N-C_L}{C_L}=\frac{C_N(X)}{C_L(X)}-1=\xi(X)-1
\end{equation}
where we have defined the drag anisotropy function $\xi(X) = C_N(X)/C_L(X)$. 

Furthermore, because the body does not deform along $x$, the laboratory and body reference frames may be related by $X=X_{0} +x$, where $X_{0}$ is the position of the anterior-most point (the tail) of the undulator along the $X$ axis. Therefore, substituting (\ref{dragFunc}) into (\ref{odeInt}) and noting that $dX_{0}/dt=dX/dt$ yields the following differential equation

\begin{equation}\label{intODEwithXi}
    \frac{dX_0}{dt} =  4\pi^2 y_0^2\int_0^1[\xi(X_{0}+x)-1]~\cos^2{2\pi(x+t)}~dx.
\end{equation}

For a given function $\xi(X)$ this may be solved to determine the motion of the body along the landscape $X_0(t)$.

For the Aubry-Andr\'e drag function 
\begin{equation}
   \xi(X)= \Delta\sin(kX) + \Gamma \sin(\beta kX)
\end{equation}

\noindent the integral on the rhs of \ref{intODEwithXi} may be integrated analytically, producing a non-linear first-order ODE that can be solved numerically. For $k=2$, it is written as

\begin{align}\label{FinalODE}
       \frac{1}{\pi^2y_0^2} \frac{dX_0}{dt}~=&~\Delta\cos{4\pi(X_0-t)}~+\\ \nonumber
       &~\frac{\Gamma}{\beta(\beta^2-4)}\times\\ \nonumber
       &~~~~\bigg(\beta(\beta+2)~\sin{\beta\pi}~\cos{\pi(2\beta X_0-4t+\beta)}\\ \nonumber
       &~~~~+ \beta(\beta-2)~\sin{\beta\pi}~\cos{\pi(2\beta X_0+4t+\beta)}\\ \nonumber
       &~~~~-(\beta+2)(\beta-2)(\sin{2\beta X_0}-\sin{2 \beta(X_0+1)}\bigg).\nonumber
\end{align}

To calculate $X_0(t)$ we numerically solved (\ref{FinalODE}) using Mathematica's numerical ODE solver. We note that for the case where $\Gamma=0$ (a single periodic drag function), (\ref{FinalODE}) is similar to the non-uniform oscillator equation, with stable fixed points at $X_0-t=0$, which correspond with uniform motion.

}

\showmatmethods{} 

\acknow{This work was funded by NSF 2310751 and DOD W911NF2110033.}

\showacknow{} 
\bibsplit[2]

\bibliography{mainbib}

\begin{thebibliography}{10}

\bibitem{berg1973bacteria}
HC Berg, RA Anderson, Bacteria swim by rotating their flagellar filaments.
\newblock {\em\protect\JournalTitle{Nature}} \textbf{245}, 380--382 (1973).

\bibitem{lauga2016bacterial}
E Lauga, Bacterial hydrodynamics.
\newblock {\em\protect\JournalTitle{Annual Review of Fluid Mechanics}} \textbf{48}, 105--130 (2016).

\bibitem{guo2008}
Z Guo, L Mahadevan, Limbless undulatory propulsion on land.
\newblock {\em\protect\JournalTitle{Proceedings of the National Academy of Sciences}} \textbf{105}, 3179--3184 (2008).

\bibitem{marvi2014sidewinding}
H Marvi, et~al., Sidewinding with minimal slip: Snake and robot ascent of sandy slopes.
\newblock {\em\protect\JournalTitle{Science}} \textbf{346}, 224--229 (2014).

\bibitem{jayne1986kinematics}
BC Jayne, Kinematics of terrestrial snake locomotion.
\newblock {\em\protect\JournalTitle{Copeia}} pp. 915--927 (1986).

\bibitem{li2009sensitive}
C Li, PB Umbanhowar, H Komsuoglu, DE Koditschek, DI Goldman, Sensitive dependence of the motion of a legged robot on granular media.
\newblock {\em\protect\JournalTitle{Proceedings of the National Academy of Sciences}} \textbf{106}, 3029--3034 (2009).

\bibitem{taylor1952analysis}
GI Taylor, Analysis of the swimming of long and narrow animals.
\newblock {\em\protect\JournalTitle{Proceedings of the Royal Society of London. Series A. Mathematical and Physical Sciences}} \textbf{214}, 158--183 (1952).

\bibitem{gray1955propulsion}
J Gray, G Hancock, The propulsion of sea-urchin spermatozoa.
\newblock {\em\protect\JournalTitle{Journal of Experimental Biology}} \textbf{32}, 802--814 (1955).

\bibitem{shapere1989geometry}
A Shapere, F Wilczek, Geometry of self-propulsion at low reynolds number.
\newblock {\em\protect\JournalTitle{Journal of Fluid Mechanics}} \textbf{198}, 557--585 (1989).

\bibitem{lauga2009hydrodynamics}
E Lauga, TR Powers, The hydrodynamics of swimming microorganisms.
\newblock {\em\protect\JournalTitle{Reports on progress in physics}} \textbf{72}, 096601 (2009).

\bibitem{Zhang2014-fi}
T Zhang, DI Goldman, The effectiveness of resistive force theory in granular locomotion.
\newblock {\em\protect\JournalTitle{Phys. Fluids}} \textbf{26}, 101308 (2014).

\bibitem{Rieser2024}
JM Rieser, et~al., Geometric phase predicts locomotion performance in undulating living systems across scales.
\newblock {\em\protect\JournalTitle{Proc. Natl. Acad. Sci. U. S. A.}} \textbf{121}, e2320517121 (2024).

\bibitem{murphy2017disaster}
RR Murphy, {\em Disaster robotics}.
\newblock (MIT press), (2017).

\bibitem{maimone2006autonomous}
M Maimone, A Johnson, Y Cheng, R Willson, L Matthies, Autonomous navigation results from the mars exploration rover (mer) mission in {\em Experimental Robotics IX: The 9th International Symposium on Experimental Robotics}.
\newblock (Springer), pp. 3--13 (2006).

\bibitem{shrivastava2020material}
S Shrivastava, et~al., Material remodeling and unconventional gaits facilitate locomotion of a robophysical rover over granular terrain.
\newblock {\em\protect\JournalTitle{Science robotics}} \textbf{5}, eaba3499 (2020).

\bibitem{bechar2016agricultural}
A Bechar, C Vigneault, Agricultural robots for field operations: Concepts and components.
\newblock {\em\protect\JournalTitle{Biosystems engineering}} \textbf{149}, 94--111 (2016).

\bibitem{gcr}
I Ground Control~Robotics, Ground control robotics, inc.. (year?).

\bibitem{park2008enhanced}
S Park, et~al., Enhanced caenorhabditis elegans locomotion in a structured microfluidic environment.
\newblock {\em\protect\JournalTitle{PloS one}} \textbf{3}, e2550 (2008).

\bibitem{majmudar2012experiments}
T Majmudar, EE Keaveny, J Zhang, MJ Shelley, Experiments and theory of undulatory locomotion in a simple structured medium.
\newblock {\em\protect\JournalTitle{Journal of the Royal Society Interface}} \textbf{9}, 1809--1823 (2012).

\bibitem{rieser2019dynamics}
JM Rieser, et~al., Dynamics of scattering in undulatory active collisions.
\newblock {\em\protect\JournalTitle{Physical Review E}} \textbf{99}, 022606 (2019).

\bibitem{schiebel2019mechanical}
PE Schiebel, et~al., Mechanical diffraction reveals the role of passive dynamics in a slithering snake.
\newblock {\em\protect\JournalTitle{Proceedings of the National Academy of Sciences}} \textbf{116}, 4798--4803 (2019).

\bibitem{Wang2023mechanical}
T Wang, et~al., Mechanical intelligence simplifies control in terrestrial limbless locomotion.
\newblock {\em\protect\JournalTitle{Science Robotics}} \textbf{8}, eadi2243 (2023).

\bibitem{transeth2008snake}
AA Transeth, RI Leine, C Glocker, KY Pettersen, P Liljeb{\"a}ck, Snake robot obstacle-aided locomotion: Modeling, simulations, and experiments.
\newblock {\em\protect\JournalTitle{IEEE Transactions on Robotics}} \textbf{24}, 88--104 (2008).

\bibitem{wang2020directional}
T Wang, J Whitman, M Travers, H Choset, Directional compliance in obstacle-aided navigation for snake robots in {\em 2020 American Control Conference (ACC)}.
\newblock (IEEE), pp. 2458--2463 (2020).

\bibitem{hirose1993biologically}
S Hirose, Biologically inspired robots.
\newblock {\em\protect\JournalTitle{Snake-Like Locomotors and Manipulators}} (1993).

\bibitem{coppeliaSim}
E Rohmer, SPN Singh, M Freese, Coppeliasim (formerly v-rep): a versatile and scalable robot simulation framework in {\em Proc. of The International Conference on Intelligent Robots and Systems (IROS)}.
\newblock (2013).

\bibitem{hu2009mechanics}
DL Hu, J Nirody, T Scott, MJ Shelley, The mechanics of slithering locomotion.
\newblock {\em\protect\JournalTitle{Proceedings of the National Academy of Sciences}} \textbf{106}, 10081--10085 (2009).

\bibitem{aubry1980analyticity}
S Aubry, G Andr{\'e}, Analyticity breaking and anderson localization in incommensurate lattices.
\newblock {\em\protect\JournalTitle{Ann. Israel Phys. Soc}} \textbf{3}, 133 (1980).

\bibitem{dominguez2019aubry}
G Dom{\'\i}nguez-Castro, R Paredes, The aubry--andr{\'e} model as a hobbyhorse for understanding the localization phenomenon.
\newblock {\em\protect\JournalTitle{European Journal of Physics}} \textbf{40}, 045403 (2019).

\bibitem{anderson1958absence}
PW Anderson, Absence of diffusion in certain random lattices.
\newblock {\em\protect\JournalTitle{Physical review}} \textbf{109}, 1492 (1958).

\bibitem{Modugno2009}
G Modugno, Exponential localization in one-dimensional quasi-periodic optical latticess.
\newblock {\em\protect\JournalTitle{New Journal of Physics}} \textbf{11}, 033023 (2009).

\bibitem{Albert2010}
M Albert, P Leboeuf, Localization by bichromatic potentials versus anderson localization.
\newblock {\em\protect\JournalTitle{Physical Review A}} \textbf{81}, 013614 (2010).

\bibitem{Zippelius2019}
Z Mokhtarin, A Zippelius, Dynamics of active filamenents in porous media.
\newblock {\em\protect\JournalTitle{Physical Review Letters}} \textbf{123}, 028001 (2019).

\bibitem{Roati2008anderson}
G Roati, et~al., Anderson localization of a non-interacting bose-einstein condensate.
\newblock {\em\protect\JournalTitle{Nature}} \textbf{453}, 895--898 (2008).

\bibitem{AspectInguscio2009}
A Aspect, M Inguscio, Anderson localization of ultracold atoms.
\newblock {\em\protect\JournalTitle{Physics Today}} \textbf{62}, 30 (2009).

\bibitem{schiebel2020mitigating}
PE Schiebel, et~al., Mitigating memory effects during undulatory locomotion on hysteretic materials.
\newblock {\em\protect\JournalTitle{Elife}} \textbf{9}, e51412 (2020).

\bibitem{rieser2021functional}
JM Rieser, TD Li, JL Tingle, DI Goldman, JR Mendelson~III, Functional consequences of convergently evolved microscopic skin features on snake locomotion.
\newblock {\em\protect\JournalTitle{Proceedings of the National Academy of Sciences}} \textbf{118}, e2018264118 (2021).

\bibitem{rabets2014direct}
Y Rabets, M Backholm, K Dalnoki-Veress, WS Ryu, Direct measurements of drag forces in c. elegans crawling locomotion.
\newblock {\em\protect\JournalTitle{Biophysical journal}} \textbf{107}, 1980--1987 (2014).

\bibitem{diaz2022water}
K Diaz, B Chong, S Tarr, E Erickson, DI Goldman, Water surface swimming dynamics in lightweight centipedes.
\newblock {\em\protect\JournalTitle{arXiv preprint arXiv:2210.09570}} (2022).

\bibitem{zhang2021friction}
X Zhang, N Naughton, T Parthasarathy, M Gazzola, Friction modulation in limbless, three-dimensional gaits and heterogeneous terrains.
\newblock {\em\protect\JournalTitle{Nature communications}} \textbf{12}, 6076 (2021).

\bibitem{pybullet}
E Coumans, Y Bai, Pybullet, a python module for physics simulation for games, robotics and machine learning (\url{http://pybullet.org}) (2016--2021).

\end{thebibliography}

\end{document}